\documentclass[%
reprint,
superscriptaddress,
amsmath,amssymb,
aps,
floatfix,
]{revtex4-2}

\usepackage{graphicx}% Include figure files
\usepackage{dcolumn}% Align table columns on decimal point
\usepackage{bm}% bold math
\usepackage{hyperref}% add hypertext capabilities
\usepackage[dvipsnames]{xcolor}
\usepackage{color}
\hypersetup{colorlinks = true,
urlcolor  = black, 
linkcolor = NavyBlue, 
citecolor = ForestGreen}
\newcommand{\sectitle}[1]{\textbf{\em #1 ---}}

\newcommand{\nE}{\mathrm{e}}
\newcommand{\D}{\mathrm{d}}
\newcommand{\A}{\langle \mathcal{A} \rangle_\tau}
\newcommand{\AD}{\langle \mathcal{A}^\mathrm{D} \rangle_\kappa}
\newcommand{\PS}{\mathrm{PS}}

\newcommand{\eqL}[1]{\label{eq:#1}}
\newcommand{\eqR}[1]{Eq.~(#1)}

\newcommand{\figL}[1]{\label{fig:#1}}
\newcommand{\figR}[1]{Fig.~(#1)}

\begin{document}

\title{General limit to thermodynamic annealing performance}

\author{Yutong Luo}
\email{yutong.luo21@imperial.ac.uk}
\affiliation{Blackett Laboratory, Imperial College London, London SW7 2AZ, United Kingdom}
\affiliation{Department of Physics, Southern University of Science and Technology, Shenzhen 518055, China}

\author{Yi-Zheng Zhen}
\email{zhenyizheng@ustc.edu.cn}
\affiliation{Hefei National Research Center for Physical Sciences at the Microscale and School of Physical Sciences, University of Science and Technology of China, Hefei 230026, China}
\affiliation{Shanghai Research Center for Quantum Science and CAS Center for Excellence in Quantum Information and Quantum Physics, University of Science and Technology of China, Shanghai 201315, China}

\author{Xiangjing Liu}
\email{liuxj@mail.bnu.edu.cn}
\affiliation{Department of Physics, Southern University of Science and Technology, Shenzhen 518055, China}

\author{Daniel Ebler}
\email{ebler.daniel1@huawei.com}
\affiliation{Theory Lab, Central Research Institute, 2012 Labs, Huawei Technology Co. Ltd., Hong Kong Science Park, Hong Kong SAR, China}
\affiliation{Department of Computer Science, The University of Hong Kong, Pokfulam Road, Hong Kong SAR, China}

\author{Oscar Dahlsten}
\email{oscar.dahlsten@cityu.edu.hk}
\affiliation{Department of Physics, City University of Hong Kong, Tat Chee Avenue, Kowloon, Hong Kong SAR, China}
\affiliation{Shenzhen Institute for Quantum Science and Engineering and Department of Physics, Southern University of Science and Technology, Shenzhen 518055, China}
\affiliation{Institute of Nanoscience and Applications, Southern University of Science and Technology, Shenzhen 518055, China}

\date{\today}

\begin{abstract}
Annealing has proven highly successful in finding minima in a cost landscape. Yet, depending on the landscape, systems often converge towards local minima rather than global ones. In this Letter, we analyse the conditions for which annealing is approximately successful in finite time. We connect annealing to stochastic thermodynamics to derive a general bound on the distance between the system state at the end of the annealing and the ground state of the landscape. This distance depends on the amount of state updates of the system and the accumulation of non-equilibrium energy, two protocol and energy landscape dependent quantities which we show are in a trade-off relation. We describe how to bound the two quantities both analytically and physically. This offers a general approach to assess the performance of annealing from accessible parameters, both for simulated and physical implementations. 
\end{abstract}

\maketitle

\noindent\sectitle{Introduction}
Annealing is the process of reaching desired physical and chemical properties by gradual external temperature control ~\cite{callister2018materials}. Systematic cooling allows the physical system to transition towards energetically more favourable states--see ~\figR{\ref{fig:curve}}. This process inspired heuristic algorithms, known as simulated annealing (SA) ~\cite{Kirkpatrick1983,Bertsimas1993,Ingber1993,salamon2002facts}, to solve optimization problems for a large variety of scientific disciplines, such as logistics, manufacturing, computer vision, machine learning and bio-informatics~\cite{BERNARDI2015872, EKREN2010592, MEIRI2006842, Mohseni2022}. 
\begin{figure}[ht]
    \centering
    \includegraphics[width=\columnwidth]{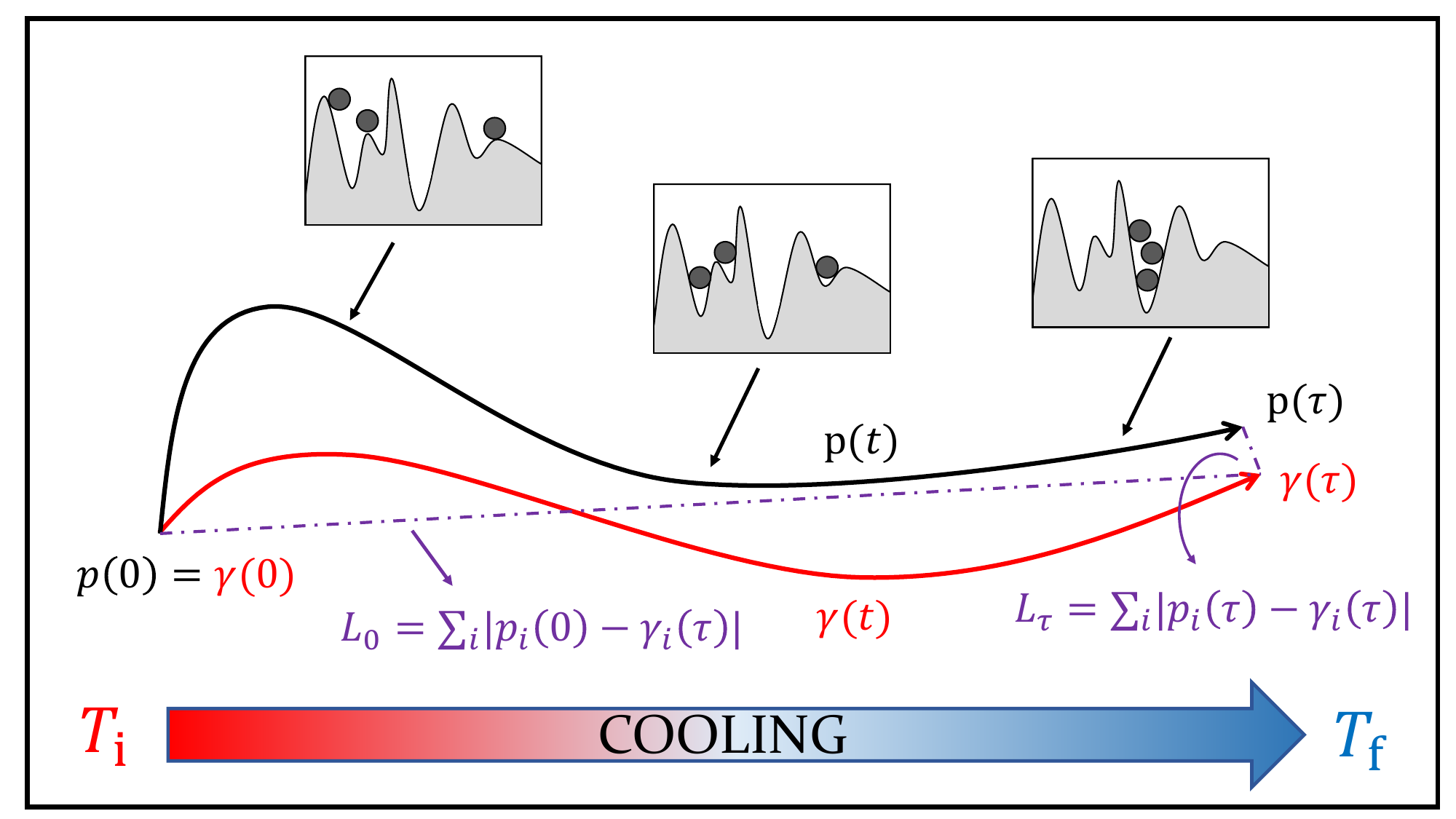}
    \caption{{\bf Annealing.} The system is initially in thermal equilibrium with a heat bath at temperature $T_\mathrm{i}$. As the temperature of the bath decreases to $T_\mathrm{f}$ during a time $\tau$, the statistical state of the system $p(t)$ (upper solid black curve) lags behind the instantaneous thermal state $\gamma(t)$ (lower solid red curve). The distances to the final thermal state $\gamma(\tau)$ for $p(0)$ and $p(\tau)$ are defined as $L_0$ and $L_\tau$ (dottdashed purple lines), respectively, with $L_\tau$ quantifying the performance error of the annealing. 
    The insets illustrate possible microstates (black balls) within the energy landscape.
    }
    \figL{curve}
\end{figure}
Simulated annealing identifies minima in optimization landscapes through stochastic updates of the system state in accordance with the energy difference to a randomly picked nearby candidate state~\cite{Kirkpatrick1983}. The success of finding approximately the  global minimum of the optimization landscape (ground state) depends crucially on the external control schedule of the system temperature. To end in a global minimum with probability 1,  an infinite annealing time is required~\cite{Hajek1988}. Hence, for finite-time schedules, there is {\em a priori} a risk that the output is strongly sub-optimal, such as a shallow local minimum in the energy landscape. These issues are not restricted to simulated annealing, appearing in physical implementations and quantum annealing~\cite{Cai2020,Mohseni2022}.

There has been significant progress in estimating the performance of annealing in finite time. 

Refs.~\cite{Hajek1988,Mitra1985,Nolte2000,Fontenas2003,Albrecht2004} obtained similar bounds on the departure from the optimal probability distribution over states. The tightness of such a bound was successfully demonstrated in a specific problem~\cite{Albrecht2004} and a bound was derived for the case of solving a graph colouring problem for general cooling schedules~\cite{Nolte1996}. Nevertheless, there are significant restrictions remaining in our understanding of the convergence towards the desired thermal state in finite time.
The bounds in Refs.~\cite{Hajek1988,Mitra1985,Nolte2000,Fontenas2003,Albrecht2004} are restricted to the case of the temperature varying as the inverse of the logarithm of the time. Some reported bounds moreover require knowledge of the system's statistical state at an intermediate time during the annealing~\cite{Nolte2000,Fontenas2003}.  The previous results nevertheless give hope that it may be possible to overcome these challenges and bound the performance of annealing more generally, as well as in terms of accessible parameters rather than the statistical state of the system.

In this Letter, we address the aforementioned challenges and derive an analytical bound on the performance of annealing. We employ recent methods from finite-time stochastic thermodynamics~\cite{Gupta2012,Esposito2010,Esposito2011, Shiraishi2018,Dechant2021,Tan2022} and generalise a speed-limit related to entropy production ~\cite{Shiraishi2018,Dechant2021,Tan2022}. This allows us to bound the distance of the final system state to the target state (which is denoted as $L_\tau$ as depicted in \figR{\ref{fig:curve}})  in terms of the protocol time $\tau$. Employing concepts from stochastic thermodynamics results in a bound that can be evaluated in terms of macroscopic thermodynamical quantities such as heat transfer. 

The bound applies very generally. It holds for all cooling schedules and only depends on two case-specific quantities. These are related to the number of state transitions of the system and the energy difference to the equilibrium energy respectively. We show that these two quantities are in a trade-off relation. 

To illustrate how the bound can be applied, including how those two quantities can be evaluated without knowing the intermediate statistical states, we apply the bound to a simulated annealing algorithm and to a physical annealing process. 

Our results thus link recent results from finite time stochastic thermodynamics with the problem of annealing, giving a performance guarantee for annealing based on thermodynamical parameters. Full details of the derivation and further studies into the application of the results to simulated annealing are provided in an accompanying paper~\cite{long}.

\noindent\sectitle{Stochastic thermodynamics of annealing} 
We first introduce stochastic thermodynamics (as used e.g.\ in Refs.~\cite{Esposito2010,Esposito2011,Dechant2021}) as the model for annealing. The anneal process puts a system into contact with a heat bath of time-dependent temperature $T(t)$. The system has $N$ energy levels denoted as $\{E_i\}_{i=1}^N$, and the corresponding probabilities of the system to be in $E_i$ at time $t$ is denoted by $p_i(t)$. The vector $p(t) = \left[p_1(t),\dots,p_N(t)\right]$ is then the {\em statistical} state of the system. A fully thermalized system is modelled to be in the thermal Gibbs' state $\gamma = \left[\gamma_i,\dots,\gamma_N\right]$ with $\gamma_i = \exp\left(-\beta E_i\right)/Z$, where $Z = \sum_i\exp\left(-\beta E_i\right)$ is the partition function and $\beta = 1/T$ is the (time-dependent) inverse temperature (The Boltzmann constant $k_\mathrm{B}$ is taken to be 1).

The system evolution under the annealing process is modelled as a master equation~\cite{Kampen2007,Esposito2010,levin2017markov}: $\dot{p}_i(t) = \sum_j\Gamma_{ij}(t)p_j(t)$, where $\Gamma_{ij}(t)$ is the generator satisfying $\sum_i\Gamma_{ij}(t) = 0, \forall j$ and $\Gamma_{ij}(t) \ge 0, \forall i\neq j$. We assume the process obeys the detailed balance condition, $\Gamma_{ij}(t)\gamma_j(t) = \Gamma_{ji}(t)\gamma_i(t)$, $\forall i,j$, as commonly assumed in non-equilibrium thermodynamics (see e.g.\ Refs.~\cite{Esposito2011, Shiraishi2018,Zhen2021}).

In non-quasistatic annealing, the state of the system $p(t)$ chases after the instantaneous thermal state $\gamma(t)$ but fails to catch up, as is shown in~\figR{\ref{fig:curve}}. To characterize the performance of an annealing protocol run in finite time $\tau$, we employ the commonly used 1-norm distance to the final thermal state $\gamma(\tau)$:  
\begin{equation}
    L_t := \sum_i|p_i(t)-\gamma_i(\tau)|.
    \label{eq:L_t}
\end{equation} For protocols with $T(\tau) = 0$, the probability of not having the ground state, $p(\text{non-optimal})$, respects $\frac{1}{2}L_\tau\geq p(\text{non-optimal})$~\cite{long}.
We will bound the protocol performance error $L_\tau$ in terms of $L_0$ and the protocol time $\tau$.

In quantifying how non-quasistatic a process is, a crucial quantity in stochastic thermodynamics is the \textit{entropy production rate}  $\dot{\Sigma}(t) \equiv \dot{S}_p(t) - \beta(t)\dot{Q}(t)$, where the system entropy $S_p(t) \equiv -\sum_ip_i(t)\ln p_i(t)$, such that $\dot{S}_p(t)$ is the rate of change of system entropy and $\dot{Q}(t) \equiv \sum_i\dot{p}_i(t)E_i(t)$ is the rate of heat transferred into the system (see e.g.\ Ref.~\cite{Esposito2011}). One can show that for processes respecting detailed balance, $\dot{\Sigma}(t)\geq 0$ with equality for quasi-static processes.

\noindent \sectitle{Universal bound on annealing performance} 
We now describe the universal bound and its derivation.
The derivation of our bound tracks the evolution of a thermodynamically natural measure of the (time-varying) difference between the statistical state and the thermal state: the relative entropy  $S(p(t)||\gamma(t)) = \sum_i p_i(t)\ln\left(p_i(t)/\gamma_i(t)\right)$. $S(p(t)||\gamma(t))$ has a thermodynamical meaning as the difference between the system's free energy and the free energy of its thermal state (see e.g.\ Ref.~\cite{Zhen2021}).

During the annealing,
\begin{equation}\label{eq:relentropyrate2parts}
\dfrac{\mathrm{d}}{\mathrm{d}t}S(p||\gamma) = \dot{p}\frac{\partial } {\partial p}S(p||\gamma)+\dot{\beta}\frac{\partial}{ \partial \beta}S(p||\gamma),
\end{equation}
since the probabilities $p(t)$ change (according to the master equation) and $\gamma(t)$ changes (solely) due to the varying temperature. One can verify that $\dot{p}\partial_p S(p||\gamma) =-\dot{\Sigma}$ (see e.g.\ Ref.~\cite{Zhen2021}), whereas $\dot{\beta}{\partial_{\beta}}S(p||\gamma):=\dot{\mathcal{I}}$, {\em the rate of change of relative entropy due to the variation in temperature}, is a novel annealing-related quantity.

Integrating \eqR{\ref{eq:relentropyrate2parts}} from time 0 (for which the system is in a thermal state) to time $\tau$ to find the final difference between the statistical and equilibrium states yields
\begin{equation}\eqL{relative_entropy}
    S(p(\tau)||\gamma(\tau)) = -\Sigma(\tau) + \mathcal{I}(\tau),
\end{equation}
where the accumulated change in relative entropy due to varying temperature 
\begin{equation}\eqL{I_def}
    \mathcal{I}(\tau) \equiv \int_0^\tau \left[E_p(t) - E_\gamma(t)\right]\dot{\beta}(t)\D t. 
\end{equation}
$E_p(t) \equiv \sum_i p_i(t)E_i$ and $E_\gamma(t) \equiv \sum_i \gamma_i(t)E_i$ denote energy expectation values. We shall later use \eqR{\ref{eq:I_def}} to give $\mathcal{I}(\tau)$ a more practical operational meaning via calorimetry.

To bound the entropy production $\Sigma(\tau)$ in \eqR{\ref{eq:relative_entropy}}, we adopt the \textit{speed limit} proposed in~\cite{Shiraishi2018}, which shows that the speed of state transformation in a thermal process is limited by the entropy production. This speed limit was originally proved in the fixed-temperature scenario and we here extend it to the varying-temperature scenario with full details in~\cite{long}.
\begin{equation}
    \Sigma(\tau) \ge \dfrac{\left(\sum_i|p_i(0)-p_i(\tau)|\right)^2}{2\A \tau} \ge \dfrac{\left(L_0-L_\tau\right)^2}{2\A \tau}, 
    \eqL{speed_limit}
\end{equation}
where $\A=\frac{1}{\tau}\int_0^\tau dt \mathcal{A}(t)$ is the time-averaged activity with the activity $\mathcal{A}(t)$ defined as \cite{Shiraishi2018,Dechant2021,Tan2022} 
\begin{equation}
    \mathcal{A}(t) \equiv \sum_i\sum_{j(\neq i)}\Gamma_{ij}(t)p_j(t).
    \eqL{activity_def}
\end{equation}
The activity can be shown to equate to the time-averaged expected number of jumps between states: ${\langle N_\text{jumps}\rangle}/{\tau} =\A$.

To bound the performance of annealing $L_\tau$, we substitute \eqR{\ref{eq:speed_limit}} and Pinsker's inequality \cite{pinsker1964information}, that $L_\tau^2\le 2S(p(\tau)||\gamma(\tau))$, into \eqR{\ref{eq:relative_entropy}}. This yields (see full derivations in~\cite{long})
\begin{equation}\eqL{main_bound}
    L_\tau \le \dfrac{L_0 + \sqrt{\langle\mathcal{A}\rangle_\tau\tau\left[-L_0^2 + 2\mathcal{I}(\tau)\left(\langle\mathcal{A}\rangle_\tau\tau + 1\right)\right]}}{\langle\mathcal{A}\rangle_\tau\tau+1}.
\end{equation}
The term $L_0$ in \eqR{\ref{eq:main_bound}} is the 1-norm distance between the {\em initial} state and the ideal final state. $L_0$ is, by inspection, constant if the initial and final temperatures ($T_\mathrm{i}$ and $T_\mathrm{f}$) are fixed, which is normally the case in annealing. $L_0$ moreover forces, as can be shown from \eqR{\ref{eq:main_bound}}, a trade-off relation between the activity $\A$ and the relative entropy change from temperature variation $\mathcal{I}(\tau)$:  $\mathcal{I}(\tau)\left(\A\tau+1\right) \ge {L_0^2}/{2}$.
The trade-off relation shows for example that $\mathcal{I}(\tau)$ can only be small if the expected number of jumps between different states $\langle N_\mathrm{jumps} \rangle\equiv\A\tau$ is large. With this relation, in the accompanying paper~\cite{long} we show that the bound scales as $O(\tau^{-\alpha/2})$ with $0<\alpha\le 1$, for protocols that converge to quasistatic processes as $\tau\rightarrow \infty$. The bound is hence tight in the quasistatic limit for any models. 

Apart from $L_0$, evaluating the general bound of~\eqR{\ref{eq:main_bound}} involves the relative entropy change from temperature variation $\mathcal{I} (\tau)$ and the time-averaged activity $\langle \mathcal{A}\rangle_\tau$. We therefore now analyse how these can be evaluated in specific protocols. 

\noindent\sectitle{Annealing an arbitrary two-level system}
To illustrate and motivate a technique we shall use for applying the general bound of~\eqR{\ref{eq:main_bound}} to specific examples, we consider an arbitrary two-level system undergoing an annealing process. The energy levels are labelled by $0$ and $1$, corresponding to energies $E_0=0$ and $E_1=E>0$, respectively. The master equation of any thermalising state evolution of the two-level system is known to be equivalent to $\dot{p_i}(t) = -\mu(t)\left(p_i(t)-\gamma_i(t)\right)$, where $i\in\{0,1\}$ and $\mu(t) = \Gamma_{01}(t) + \Gamma_{10}(t)$ is the {\em partial swap} (PS) rate~\cite{Scarani2002,Browne2014,Zhen2021}, so called since the model is equivalent to swapping the state with a thermal state $\gamma$ with probability $\mu dt$ for time interval $dt$. The bound in \eqR{\ref{eq:main_bound}} here becomes
\begin{equation}
    L_\tau \le \dfrac{L_0 + \sqrt{2\langle\mu\rangle_\tau\tau\left[-L_0^2 + 2\mathcal{I}^\PS(\tau)\left(2\langle\mu\rangle_\tau\tau + 1\right)\right]}}{2\langle\mu\rangle_\tau\tau+1},
\end{equation}
where $\langle\mu\rangle_\tau =  \tau^{-1} \int^\tau_0\mu(t)\D t$ and 
\begin{equation}
    \mathcal{I}^\PS(\tau) = -\int_0^\tau\int_0^t\nE^{-\int_{s}^t\mu(s')\D s'}\dot{E}_\gamma(s)\dot{\beta}(t)\D s\D t,
    \eqL{I_PS_tau}
\end{equation}
with $E_\gamma(t) = E / \left(1+\nE^{\beta(t)E}\right)$ (see Sec.~I of Supplemental Material~\footnote{See Supplemental Material at \url{https://journals.aps.org/pre/supplemental/10.1103/PhysRevE.108.L052105/SM.pdf} for the description of the partial swap model, a lower bound for $L_t$, and more numerical results for SK models.} for details).

Importantly, \eqR{\ref{eq:I_PS_tau}} can be evaluated without the system's energy $E_p(t)$. We shall exploit this property of the partial swap model, which holds for any number of energy levels, to bound
$\mathcal{I}(\tau)$ in other models via a suitable choice of partial swap rate $\mu$.

\noindent\sectitle{Bound for SA}
We now derive how to evaluate the performance bound for Simulated Annealing (SA). In SA the time evolution obeys detailed balance and is discrete, taking  place in $\kappa$ discrete time steps. We, as will be described below with full details in \cite{long}, modify our arguments to discrete time and bound $\mathcal{I}$ and $\mathcal{A}$, finding 
\begin{equation}\eqL{SA_bound}
    L_\kappa \le \dfrac{2L_0 + \sqrt{2\kappa\left[-L_0^2 + \mathcal{I}^\PS(\kappa)\left(\kappa + 2\right)\right]}}{\kappa+2}.
\end{equation}
$\mathcal{I}^\PS(\kappa)$ is the discrete-time relative entropy change from the variation of temperature for the partial swap model of thermalisation, a model discussed in the previous section. For \eqR{\ref{eq:SA_bound}} to hold, the choice we make for the partial swap rate is 
\begin{equation}\eqL{mu_PS}
 \mu^\PS(k) = \dfrac{1}{n}\exp\left[-\beta(k+1)\Delta E_{\max}\right],
\end{equation}
where $n$ is the number of accessible states within each SA step, $\Delta E_{\max} = \max_iE_i - \min_iE_i$. This $\mu^\PS(k)$ is chosen to ensure the partial swap model relaxation time is not faster than a known bound~\cite{Desai1993} for SA models~\cite{long}. 
 $\mathcal{I}^\PS(\kappa) = \sum_{k=1}^{\kappa}[E_p^\PS(k)-E_\gamma(k)][\beta(k)-\beta(k-1)]$, where $E_p^\PS(k)$ is defined iteratively according to  $ E_p^\PS(k+1) = [1-\mu^\PS(k)]E_p^\PS(k) + \mu^\PS(k)E_\gamma(k+1)$.

\eqR{\ref{eq:SA_bound}} has two remarkable features. Firstly, it is history-independent, in the sense that $\mathcal{I}^\PS(\kappa)$ can be evaluated only by the equilibrium energy $E_\gamma(k)$. The fact that the intermediate statistical state $p(k)$ is then no longer required heavily reduces the computational cost of evaluating the statistical error $L_{\kappa}$~\cite{long}. Secondly, \eqR{\ref{eq:SA_bound}} holds for all cooling schedules $\beta(k)$.

The derivation of \eqR{\ref{eq:SA_bound}} can be broken into three steps. 
Firstly, we assume a particular form of dynamics satisfying detailed balance associated with SA: Glauber dynamics~\cite{Glauber1963,Levin2008,Walter2015}. We show that the discrete-time analogy of the speed limit from entropy production
% the entropy production speed limit for varying temperatures 
[\eqR{\ref{eq:speed_limit}}] here becomes {$\Sigma(\kappa) \ge (L_0-L_\kappa)^2/(2\langle \mathcal{A}\rangle_\kappa\kappa) \ge (L_0-L_\kappa)^2/\kappa$}, which rests on the restriction that the (discrete-time) activity $\mathcal{A}^\mathrm{D}(k) \le 1/2$, $\forall k$. 
One can show that for the infinite temperature thermal state $\mathcal{A}^\mathrm{D} = 1/2$ and for any lower temperature thermal states $\mathcal{A}^\mathrm{D} < 1/2$ with $\mathcal{A}^\mathrm{D}=0$ for zero temperature thermal states. We conjecture, based on those and further analytical arguments and numerics that in general $\mathcal{A}^\mathrm{D}\le 1/2$ for SA with Glauber dynamics, which consequently gives $\AD\le 1/2$ (see Ref.\cite{long} for details).

We then derive, up to a conjecture, that $\mathcal{I}(\kappa) \leq \mathcal{I}^\PS(\kappa)$. We show that the relaxation time~\cite{levin2017markov} of the actual process can be bounded by that of a suitably chosen partial swap model: $\tau_\mathrm{rel}(k)\le\tau_\mathrm{rel}^\PS(k)$. The conjecture is then that $\tau_\mathrm{rel}(k)\le\tau_\mathrm{rel}^\PS(k) \Rightarrow \mu(k)\ge\mu^\PS(k)$, where $\mu(k)$ is a particular rate associated with the actual process and $\mu^\PS(k)$ is defined in \eqR{\ref{eq:mu_PS}}. From $\mu(k)\ge\mu^\PS(k)$ one can prove $\mathcal{I}(\kappa) \leq \mathcal{I}^\PS(\kappa)$ (see Ref.\cite{long} for details). 

Finally, as in the derivation of \eqR{\ref{eq:main_bound}}, we combine the above bounds on $\Sigma (\kappa)$ and $\mathcal{I}(\kappa)$, Pinsker's inequality
and \eqR{\ref{eq:relative_entropy}}, to obtain \eqR{\ref{eq:SA_bound}}.

\noindent\sectitle{Tightness of bound}
To show the tightness of the SA version of the bound [\eqR{\ref{eq:SA_bound}}] we present, in \figR{\ref{fig:result_SK}}, simulation results on a $7$-spin Sherrington-Kirkpatrick (SK) model, which is a fully-connected spin glass with Gaussian couplings~\cite{SKmodel,Panchenko2013}. 
The left plot of \figR{\ref{fig:result_SK}}, shows that both bounds of $L_\kappa$ outperform the constant bound $L_0$ after a few steps,  capturing the relaxation dynamics in SA. 
The right plot verifies that $\mathcal{I}^\PS(\kappa)\ge \mathcal{I}(\kappa)$. The right plot also verifies the trade-off relation  between 
$\mathcal{I}(\kappa)$ and $\AD$, given after \eqR{\ref{eq:main_bound}}, (using $\AD\le 1/2$). One sees that the restriction on $\mathcal{I}(\kappa)$ gets tighter as $\kappa$ increases. For large time $\kappa$, $\mathcal{I}(\kappa)\sim1/\kappa$, in line with the trade-off relation. A similar agreement was found for another type of cooling schedule~\cite{long}.

Further numerical results provided in Sec.~II of Supplemental Material~\cite{Note1} show the tightness of Eqs.~(\ref{eq:main_bound}) and (\ref{eq:SA_bound}) for SK models with different sizes $n$. \eqR{\ref{eq:SA_bound}} is found to be generally tighter for the common case of high initial temperature.

\begin{figure}[htb]
    \centering
    \includegraphics[width=\columnwidth]{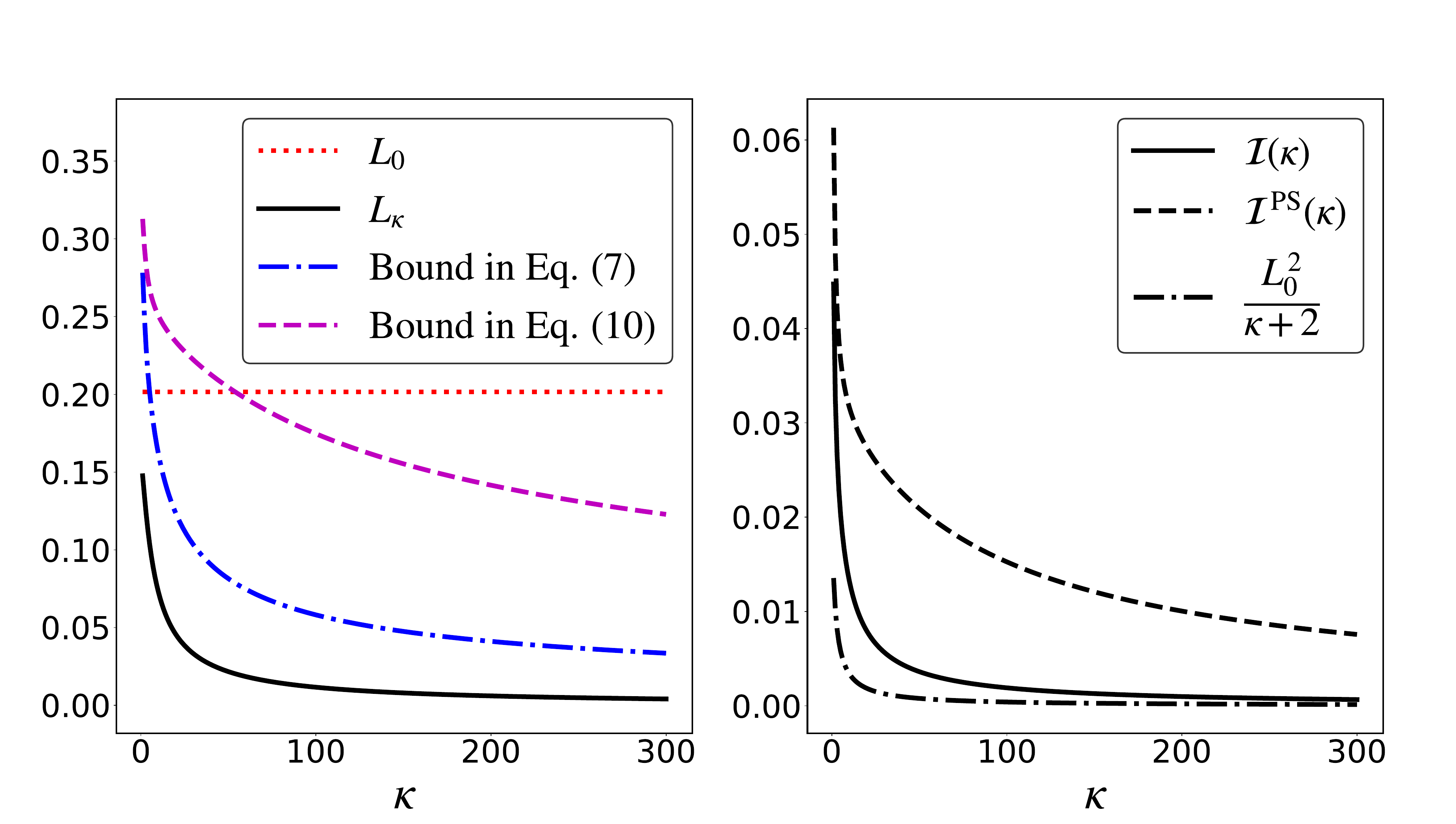}
    \caption{\figL{result_SK}{\bf Bounds on the performance of SA in solving a 7-spin SK model.} 
    $T_\mathrm{i} = 1.8$ is decreased linearly to $T_\mathrm{f} = 0.8$ in $\kappa$ steps. {\em Left:} The performance error $L_\kappa$ and our bounds in \eqR{\ref{eq:main_bound}} and in \eqR{\ref{eq:SA_bound}}, as well as the performance error of the initial state $L_0$.
    {\em Right:} The accumulated relative entropy due to temperature change $\mathcal{I}(\kappa)$ with our upper bound $\mathcal{I}^\PS(\kappa)$ and our lower bound from the trade-off relation between $\mathcal{I}(\kappa)$ and $\AD$ (for $\AD\le 1/2$).}
\end{figure}

\noindent\sectitle{Physical annealing}
In this section, we consider the possibility of implementing our results in real annealing processes. By outlining a dynamic calorimetry method \cite{Garden2007}, we will show that the energy difference $E_p(t)-E_\gamma(t)$ can be measured during the annealing process. Thanks to this technique,
the generally unknown distance $L_t$ (\eqR{\ref{eq:L_t}}) can be bounded by (see Sec.~III of Supplemental Material~\cite{Note1} for details)
\begin{equation}
    L_t\ge \tilde{L}_t := \frac{2\left[E_p(t)-E_\gamma(\tau)\right]}{\Delta E_{\max}},
\end{equation}
where $\Delta E_{\max} = \max_i E_i - \min_i E_i$ as defined previously is the energy scale of the system.
The annealing performance error $L_\tau$ is thus bounded from below and above as
\begin{equation}
    \tilde{L}_\tau \le L_\tau \le \dfrac{\tilde{L}_0 + \sqrt{-\tilde{L}_0^2 + \langle\mathcal{A}\rangle_\tau\tau\left[2\mathcal{I}(\tau)\left(\langle\mathcal{A}\rangle_\tau\tau + 1\right)\right]}}{\langle\mathcal{A}\rangle_\tau\tau+1}.
    \eqL{real_bound}
\end{equation}

\begin{figure}[hbt]
    \centering
    \includegraphics[width=\columnwidth]{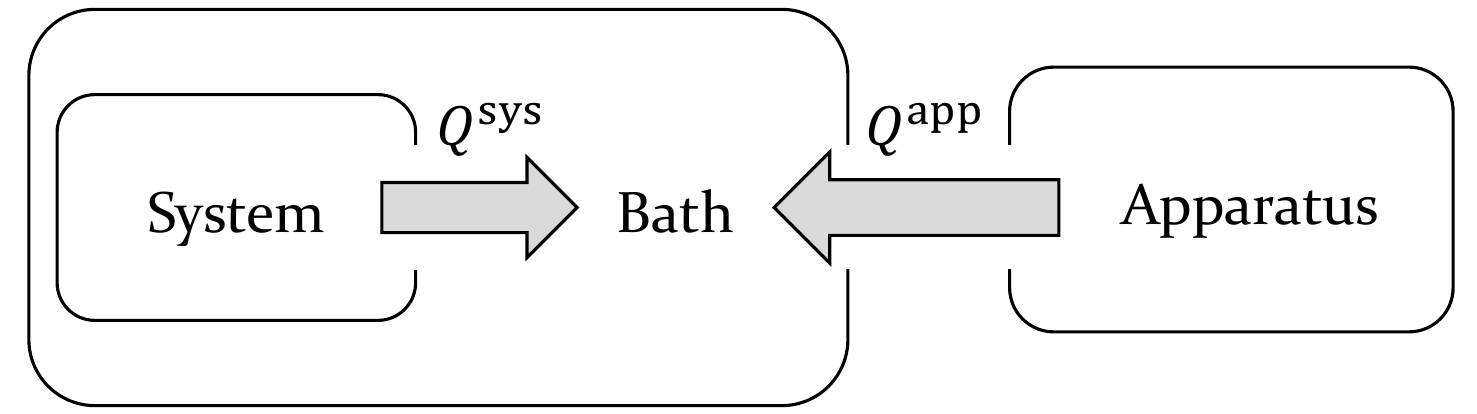}
    \caption{\textbf{Calorimetric setup}. The system in contact with the heat bath undergoes an annealing process as the bath temperature is tuned downwards. The external apparatus is accessible to supply a specific amount of heat to the bath to control the temperature of the bath and thus to apply different cooling schedules. The bath is assumed to be in thermal equilibrium all the time.}
    \figL{calorimetry}
\end{figure}

In order to measure the quantities in \eqR{\ref{eq:real_bound}}, we propose a calorimetric setup for the annealing process [see \figR{\ref{fig:calorimetry}}] where the system is connected to a heat bath whose temperature is controlled by exchanging heat with an external apparatus. The bath is assumed to be in thermal equilibrium at all times, such that its heat capacity $C^\mathrm{bath}(T)$ is well defined. The energy change of the bath is given by $  \D E_\gamma^\mathrm{bath} = \D Q^\mathrm{sys} + \D Q^\mathrm{app} = C^\mathrm{bath}(T) \D T^\mathrm{bath}$, where $\D Q^\mathrm{sys}$ and $\D Q^\mathrm{app}$ are the heat transferred to the bath from the system and the apparatus, respectively. The control of $Q^\mathrm{app}$ allows us to adjust the bath temperature to realize different cooling schedules $T^\mathrm{bath}(t)$. By energy conservation, $\D E_p = -\D Q^\mathrm{sys}$, which gives $ \D E_p = \D Q^\mathrm{app} - C^\mathrm{bath}(T) \D T^\mathrm{bath}$. Referring to the heat capacity of the system $C^\mathrm{sys}(T)$, we can also calculate the change in the equilibrium energy $\D E_\gamma = C^\mathrm{sys}(T) \D T^\mathrm{bath}$. Hence, integrating from time 0 to $t$, we obtain $ E_p(t) - E_\gamma(t) = Q^\mathrm{app}(t) - \int_{T_\mathrm{i}}^{T(t)} [C^\mathrm{bath}(T') + C^\mathrm{sys}(T')]\D T'$, where we have used $E_p(0) = E_\gamma(0)$. By measuring $Q^\mathrm{app}(t)$, we can calculate $\mathcal{I}(\tau)$ using \eqR{\ref{eq:I_def}}, gaining 
\begin{align}
    \mathcal{I}(\tau) &= \int_0^\tau Q^\mathrm{app}(t)\dot{\beta}(t)\D t \nonumber\\
        &\quad + \int_{T_\mathrm{i}}^{T_\mathrm{f}}\int_{T_\mathrm{i}}^{T} \frac{C^\mathrm{bath}(T') + C^\mathrm{sys}(T')}{T^2}\D T'\D T.
\end{align}
To evaluate \eqR{\ref{eq:real_bound}} via this calorimetric scheme, we moreover need $ \tilde{L}_0 = -\frac{2}{\Delta E_{\max}}\int_{T_\mathrm{i}}^{T_\mathrm{f}} C^\mathrm{sys}(T) \D T$, and $ \tilde{L}_\tau = \frac{2 Q^\mathrm{app}(\tau)}{\Delta E_{\max}} - 2\int_{T_\mathrm{i}}^{T_\mathrm{f}} \frac{C^\mathrm{bath}(T) + C^\mathrm{sys}(T)}{\Delta E_{\max}}\D T$. One way to estimate $\Delta E_{\max}$, is to posit that the energy spectrum exhibits  certain statistics, e.g.\ Wigner-Dyson statistics~\cite{d2016quantum} and thus  obtain a bound on $\Delta E_{\max}$ up to an error probability.
As for the time-averaged activity $\A$ in \eqR{\ref{eq:real_bound}}. It is conjectured that $\A$ relates to the diffusion coefficient in an overdamped Langevin equation \cite{Dechant2021,Tan2022}. Under this conjecture, $\A$ could be estimated by measuring the diffusion coefficient of the system. Taken together, the above arguments suggest it may be experimentally viable to evaluate \eqR{\ref{eq:real_bound}} to gauge the finite-time performance of physical annealing.

\noindent\sectitle{Summary and outlook}
We derived a universal bound on the distance from the system state to the final thermal state in a general annealing process. The bound captures the annealing time dependence of the annealing performance and can be applied to any system and any cooling schedule. Two evolutionary history-dependent quantities, $\A$ and $\mathcal{I}(\tau)$, appearing in the bound, were shown to follow a trade-off relation forced by the initial distance $L_0$. We applied this bound on a general two-level system where the partial swap model was adopted to remove the history dependence. We then generalised this method to derive a history-independent bound on the performance of the simulated annealing algorithm. In real annealing processes, we provided performance guarantees from both directions and outlined a calorimetric experimental setup to measure the involved parameters. 

Apart from implementing the method presented here for guaranteeing the performance of annealing in different scenarios, the approach can be generalised in several directions. We expect that other tools from non-equilibrium thermodynamics such as fluctuation theorems \cite{Jarzynski1997,Crooks1999,Esposito2009} and single-shot statistical mechanics\cite{dahlsten2011inadequacy, rio2011thermodynamic, horodecki2013fundamental, aaberg2013truly} can be adapted similarly to analyse simulated annealing, and that the approach can be generalised to variants of annealing such as parallel tempering~\cite{earl2005parallel}. 

\begin{acknowledgments}
We gratefully acknowledge valuable discussions with Alexander Yosifov, Barry Sanders, Li Xiao, and Yu Chai.
This work was supported by the National Natural Science Foundation of China (Grants No. 12050410246, No. 12005091). 
\end{acknowledgments}

%for APS Submission we should comment out bibtex and paste in output.bbl as described in 
%https://www.overleaf.com/learn/latex/Questions/The_journal_says_%22don%27t_use_BibTeX%3B_paste_the_contents_of_the_.bbl_file_into_the_.tex_file%22._How_do_I_do_this_on_Overleaf%3F

%\bibliographystyle{ieeetr}
\bibliography{references}

\end{document}